\documentstyle[preprint,eqsecnum,aps,epsbox]{revtex}

\begin{document}
\draft
\preprint{HEP-TH/9609067, KYUSHU-HET-33}
\title{Nonperturbative renormalization group \\
in a light-front three-dimensional real scalar model}
\author{Takanori Sugihara
\footnote{e-mail : taka@higgs.phys.kyushu-u.ac.jp}}
\address{Department of Physics, 
Kyushu University, Fukuoka 812-81, Japan}
\author{Masanobu Yahiro
\footnote{e-mail : yahiro@fish-u.ac.jp}}
\address{University of Fisheries, 
Shimonoseki 759-65, Japan}
\date{\today}
\maketitle
\thispagestyle{empty}
\setcounter{page}{0}
\begin{abstract}
The three-dimensional real scalar model, 
in which the $Z_2$ symmetry spontaneously breaks, 
is renormalized in a nonperturbative manner
based on the Tamm-Dancoff truncation of the Fock space. 
A critical line is calculated 
by diagonalizing the Hamiltonian regularized with basis functions. 
The marginal ($\phi^6$) coupling dependence 
of the critical line is weak. 
In the broken phase the canonical Hamiltonian is tachyonic, so
the field is shifted as $\phi(x)\to\varphi(x)+v$. 
The shifted value $v$ is determined 
as a function of running mass and coupling
so that the mass of the ground state vanishes.  
\end{abstract}


\vfill\eject

\section{\bf Introduction}
\label{intro}
Although the field theory has been studied for a long time, 
we have no complete understanding 
of relativistic bound state problems. 
We have to solve the quantum chromodynamics (QCD) nonperturbatively 
to investigate low-energy hadronic physics. 
The most reasonable nonperturbative method for such a problem is 
the lattice gauge theory. 
In the method, it is easy to calculate the lightest particle state, 
but not to evaluate the excited and scattering states. 
Although there exist many nonperturbative prescriptions 
in the Hamiltonian formalism, 
the formalism has been abandoned so far 
since Lorentz invariance and the renormalizability 
are not obvious. 

The light-front (LF) Tamm-Dancoff field theory 
\cite{phw} is known
as a method based on the Hamiltonian formalism. 
The Hamiltonian is constructed by quantizing fields 
on the light-cone and truncating the Fock space. 
The truncation is called 
the Tamm-Dancoff approximation \cite{tamm_dancoff}. 
The completeness of the Fock space is approximated as 
$\sum_{n=1}^{N_{\rm TD}}|n \rangle\langle n| \sim 1$, 
where $N_{\rm TD}\to\infty$ corresponds to the full theory.
It is possible to calculate mass spectra nonperturbatively 
by diagonalizing the Hamiltonian in the space.  
The resulting mass spectrum seems to be accurate 
for low-lying states, since the pair creation and annihilation 
are suppressed in the LF field theory\cite{beta}, 
especially the pair creation of particles from the LF vacuum 
is kinematically prohibited. 
The prohibition warrants the truncation to be reliable. 
This is an advantage of this method, 
but, at the same time, it causes a problem. 
No pair creation means 
that the vacuum is trivial $H|0\rangle=0$, 
i.e., the vacuum is always symmetric in the LF field theory. 
This causes serious problems, 
because various phenomena are explained 
as results of spontaneous symmetry breaking (SSB) 
in the relativistic field theory. 
How can the LF field theory describe SSB?
It has been said that the zero-mode operator 
leads the LF field theory to SSB\cite{mas_yama,kty}. 
We have a constraint equation for the zero-mode, 
if the system is quantized in a box 
with a periodic boundary condition. 
New interactions induced by the zero-mode
affect the renormalization \cite{maeno}, 
independently of whether 
the symmetry breaks spontaneously or not. 
A relation between the zero-mode scenario and 
the conventional understanding 
based on the effective potential in the equal-time 
field theory is not clarified yet. 
There are several works \cite{zero}
for the two-dimensional real scalar model,  
in which the zero-mode constraint equation 
is solved under reasonable approximations 
to describe SSB of the $Z_2$ symmetry. 
At least for the two-dimensional case, the light-front formulation  
seems to be consistent with the equal-time 
one, if the zero-mode contributions are treated properly.  
This approach to SSB has never been applied to higher dimensional 
models having ultraviolet divergences. 

According to our previous work \cite{sugi} 
based on perturbation, 
the canonical Hamiltonian yields a tachyonic spectrum 
for the broken phase owing to renormalization effects. 
We have to expand the field 
around the correct vacuum expectation value (VEV)
to restore positivity of the spectrum. 
There exists an RG invariant relation 
among VEV and running parameters, 
and the VEV is determined from the relation 
as a function of the original parameters. 
The broken phase is an RG invariant surface containing 
the critical line in the parameter space. 
The symmetry breaking interactions 
were originally introduced by Perry and Wilson 
in the coupling coherence study \cite{perry_wilson,perry}, 
but the instability of the canonical Hamiltonian was not mentioned. 

In this paper, 
we treat the real scalar model in three dimensions
and propose a nonperturbative RG prescription based on 
the Tamm-Dancoff truncation. 
The theory is regularized 
by expanding wavefunctions in terms of 
basis functions in a truncated momentum space.  
The mass spectrum is then obtained 
by diagonalizing the Hamiltonian 
within a space spanned by the basis functions. 
In actual calculations we take the Fock space 
up to three-body states. 
It is discussed through wave functions whether 
the truncated space is large enough. 
The marginal coupling dependence of the critical line 
is also analyzed. 
A proper Hamiltonian for the broken phase 
is constructed by shifting the field 
as $\phi(x)\to\varphi(x)+v$. 
A relation of $v$ to the original parameters are 
nonperturbatively determined 
by searching the critical surface 
which gives a vanishing mass for the ground state. 

The paper is organized as follows. 
In section \ref{shift}, a qualitative aspect of mass spectra 
in the three-dimensional real scalar model is shown
with perturbative Bloch-Horowitz RG analysis. 
In section \ref{basis}, 
a regularization based on basis functions is introduced. 
In section \ref{numerical}, 
numerical RG is carried out 
by using the nonperturbative regularization mentioned above. 
The section \ref{discussion} is devoted to discussions. 

Notational conventions are summarized as 
$x^{\pm}=(x^0\pm x^2)/\sqrt{2}$, \quad
time $x^+$, \quad
space $(x^-,x_\perp)$, \quad
momentum ${\bf k}\equiv(k^+,k_\perp)$, \quad
metric $g^{+-}=g_{+-}=-g^{11}=-g_{11}=1$
and others $=0$, 
and LF energy $k^- = \epsilon({\bf k}) \equiv (k_\perp^2 + r)/2k^+$, 
where $r$ is a mass parameter. 

\section{SHIFTED FIELD AND POSITIVE DEFINITE HAMILTONIAN}
\label{shift}
Let us consider a Hamiltonian 
regularized with some cut-off $\Lambda_n$
in order for a divergent theory to be well-defined. 
The Einstein-Schr\"odinger (ES) equation for the Hamiltonian is
\begin{equation}
  H(r_n,\lambda_n,\dots;\Lambda_n) | \Psi \rangle = E | \Psi \rangle, 
\end{equation}
where $r_n,\lambda_n,\dots$ are running parameters 
belonging to the model. 
We need a certain transformed Hamiltonian 
which has a lowered cutoff $\Lambda_{n+1}(\Lambda_{n+1}<\Lambda_n)$ 
and gives the same eigenvalue $E$, 
\begin{equation}
  H'(r_{n+1},\lambda_{n+1},\dots;\Lambda_{n+1}) | \Psi' \rangle
  = E | \Psi' \rangle. 
\end{equation}
We are now looking for a prescription 
to get transformed parameters $r_{n+1},\lambda_{n+1}$ which include
radiative corrections. 
To do that, we use the following projection operators\cite{lbhm}, 
\begin{equation}
  {\cal P} = \theta(\Lambda_{n+1}^2 - M_{\rm int}^2), \quad
  {\cal Q} = \theta(M_{\rm int}^2 - \Lambda_{n+1}^2), 
\end{equation}
where $M_{\rm int}$ is a total invariant mass ($M_{\rm int}^2=P^2$) 
of an intermediate Fock state. 
The transformed Hamiltonian, 
which gives a correct spectrum in the ${\cal P}$ space, 
can be written formally as, 
\begin{equation}
  \label{bloch}
  H' = {\cal P}H{\cal P} + 
  {\cal P}V{\cal Q}\frac{1}{E-H}{\cal Q}V{\cal P},
  \quad
  |\Psi' \rangle = {\cal P}|\Psi \rangle.
\end{equation}
This Hamiltonian $H'$ is called 
the Bloch-Horowitz effective Hamiltonian 
and used in many-body problem 
for the purpose of reduction of the degrees of freedom 
which are too much to treat it directly\cite{bloch_horowitz}. 
We expect that this Hamiltonian effectively
describes physics near the ground state. 
Expanding the term $1/(E-H)$ 
with respect to the coupling constant, 
integrating over the higher invariant mass space ${\cal Q}$ 
and rescaling the lowered cutoff $\Lambda_{n+1}$ 
to the original one $\Lambda_n$, 
we can write down RG equations 
for all parameters\cite{wilson_kogut}. 
The effective Hamiltonian, however, contains a problem. 
The resultant RG equations 
seems to depend on the energy $E$ at a sight. 
In our previous work\cite{sugi}, 
we have shown perturbatively 
for the four-dimensional $\phi^4$ model 
that the energy dependence of RG equation 
is negligible at one-loop order
if the cutoff is sufficiently large. 
It's not clear 
whether it is true or not beyond perturbation. 
In this paper, 
we use the effective Hamiltonian (\ref{bloch}) 
only in order to explain qualitative effects 
of RG on mass spectra perturbatively. 

Next, we derive a perturbative RG equation 
for mass parameter 
in the three-dimensional real scalar model. 
We assume that the Lagrangian density is 
\begin{equation}
  {\cal L} = \frac{1}{2}\partial_\mu \phi \partial^\mu\phi
  - \frac{r}{2}\phi^2
  - \frac{\lambda}{4!}\phi^4
  - \frac{w}{6!}\phi^6. 
\end{equation}
The field is quantized as 
\begin{equation}
  \label{com}
  [\phi(x),\phi(y)]_{x^+ = y^+}
  =-\frac{i}{4}\epsilon(x^- - y^-) \delta(x_\perp - y_\perp), 
\end{equation}
if the zero-mode is neglected. 
The factor $1/4$ on the right-hand side of Eq. (\ref{com})
is from the Poisson bracket 
\begin{equation}
  \{\Phi(x),\Phi(y)\}_{x^+ = y^+} =
  (-\partial_-^x + \partial_-^y)
  \delta(x^- - y^-)\delta(x_\perp - y_\perp), 
\end{equation}
where 
$\Phi(x)=\pi(x)-\partial_{-} \phi(x)$ 
is a primary constraint
and $\pi$ is conjugate to $\phi$. 
The canonical Hamiltonian is given by, 
\begin{eqnarray}
  \label{canno}
  H_{\rm S} = \int d^2 x 
  \left(
  \frac{1}{2} (\partial_\perp \phi)^2
  +\frac{r}{2}\phi^2
  +\frac{\lambda}{4!}\phi^4
  +\frac{w}{6!}\phi^6
  \right). 
\end{eqnarray}
The normal ordering is taken here. 
This is equivalent to renormalization of 
tad-pole diagrams by redefining parameters. 
We have to include 
various kinds of relevant and marginal interactions
to renormalize all new interactions produced by RG transformations. 
Our Hamiltonian is assumed to have three interactions
included in Eq. (\ref{canno}). 
The mass renormalization is most important, 
because $r$ is most relevant to mass spectra. 
It is the kinetic term which mainly contributes to mass spectra
especially in the perturbative region. 
The RG equation for mass $r$ is given with perturbation as 
\begin{equation}
  \label{mu}
  r_{n+1}
  = L^2
  \left( r_{n} + \frac{B}{192\pi^2}\lambda_{n}^2  \right)
  + O(\lambda_n^3), 
\end{equation}
where
\begin{equation}
  B \equiv \int \left[ \prod_{i=1}^3
  \frac{d^2 k_i}{k_i^+} \right]
  \delta^2({\bf P}-\sum_{i=1}^3{\bf k}_i)
  \left( E - \sum_{i=1}^3 \epsilon({\bf k}_i) \right)^{-1}, \quad
  L \equiv \frac{\Lambda_n}{\Lambda_{n+1}}. 
\end{equation}
It is understood implicitly that 
the loop integral is made in the ${\cal Q}$ space. 
The factor $L$ is produced by rescaling the new cutoff $\Lambda_{n+1}$ to
the original cutoff $\Lambda_n$. 
By changing variables to Jacobi coordinates, 
\begin{eqnarray*}
  &{\bf k}_i = (x_i K^+, x_i K_\perp + s_i), \quad
  \sum_{i=1}^3 x_i = 1, \quad
  \sum_{i=1}^3 s_i = 0, & \\
  \nonumber
  &0<x_i<1, \quad -\infty<s_i<\infty,&
\end{eqnarray*}
where all intermediate states have 
a common total momentum ${\bf K}$, 
the integral $B$ becomes 
\begin{equation}
  B=
  2 \int \frac{dx_1 dx_2 ds_1 ds_2}{x_1 x_2 (1- x_1 - x_2)}
  \frac{1}{M^2 - M_{\rm int}^2(x_1,x_2,s_1,s_2)}, 
\end{equation}
where $M$ is an external mass (an eigenvalue of Hamiltonian) and
$M_{\rm int}$ a mass in the intermediate Fock state,
\begin{equation}
  M_{\rm int}^2(x_1,x_2,s_1,s_2)
  \equiv
  \frac{s_1^2 + r_n}{x_1} +
  \frac{s_2^2 + r_n}{x_2} +
  \frac{(s_1 + s_2)^2 + r_n}{1-x_1-x_2}. 
  \nonumber
\end{equation}
It is assumed that
the external mass $M$ is small compared to the cutoff scale. 
The intermediate mass $M_{\rm int}$ is large 
because it sits in the higher mass space $\cal{Q}$.
Then, $M^2$ is much smaller than $M_{\rm int}^2$. 
This assumption is reasonable for the purpose 
to calculate a critical line and draw a phase diagram. 
We can observe from $B$ being negative that 
$\lim_{n\to\infty}r_n = \infty$ 
if the initial value $r_0$ is sufficiently large
and $\lim_{n\to\infty}r_n = -\infty$ 
if the initial value $r_0$ is small. 
That is, there exist a critical line $r=r_{\rm c}(\lambda)$ 
in the first quadrant
of the two dimensional parameter space $(r,\lambda)$, 
and $r>r_c(\lambda)$ and $r<r_c(\lambda)$
correspond to symmetric and asymmetric phases respectively. 
In the asymmetric phase, 
the canonical Hamiltonian (\ref{canno}) is tachyonic, i.e., 
square of eigenmass is not bounded from below
although Hamiltonian should be positive definite. 
Positivity of the spectrum is restored by 
expanding the field $\phi$ around the correct VEV $v$. 
Substituting $\phi(x)=\varphi(x)+v$ into Eq. (\ref{canno}), 
we have
\begin{eqnarray}
  \label{ham}
  H_{\rm A} &=&H_0 + H_1, \\
  H_0&=&
  \int d^2x
  \frac{1}{2}
  \left( (\partial_\perp\varphi)^2 + g_2\varphi^2  \right), \\
  H_1&=&
  \int d^2x
  \sum _{i=3}^6 \frac{g_n}{n!} \varphi^n,
\end{eqnarray}
where the new mass parameter is 
\begin{equation}
  g_2 \equiv r+\frac{\lambda v^2}{2} + \frac{wv^4}{24}, 
\end{equation}
and the new coupling constants are
\begin{equation}
  g_3 \equiv (\lambda + \frac{wv^2}{6})v, \quad
  g_4 \equiv \lambda + \frac{wv^2}{2}, \quad
  g_5 \equiv wv, \quad
  g_6 \equiv w. 
\end{equation}
The $v$ is just a free parameter at this stage. 
We have three parameters in the symmetric Hamiltonian $H_{\rm S}$
and four in the asymmetric one $H_{\rm A}$. 
The RG parameter space for $H_{\rm A}$ is larger than that for $H_{\rm S}$. 
The $v$ should be a function of the original three parameters 
belonging to $H_{\rm S}$, that is, there should  
exist an RG invariant relation among the four parameters.  
Actually, following our previous work \cite{sugi}
for the real scalar model in four 
dimensions, we can find out the same RG invariant relation 
$6r+\lambda v^2=0$ also for the three-dimensional case, 
with perturbative renormalization at one-loop order, 
if the $\phi^6$ interaction is neglected. 
The neglect of the marginal operator will be 
justified by numerical analyses shown later. 
The RG invariant relation forms 
a surface in the parameter space of $H_{\rm A}$. 
All asymmetric Hamiltonians on the surface describe physics 
for the broken phase. 
All running parameters on the surface go to zero in the limit $\Lambda\to 0$ 
which corresponds to infinite iterations of RG transformation, 
satisfying the RG invariant relation $6r+\lambda v^2=0$. 
The Hamiltonian $H_{\rm A}$ thus gives massless spectrum in the limit. 
All Hamiltonians on the surface converge to the same 
massless Hamiltonian in the limit, indicating that $H_{\rm A}$ 
gives massless spectrum at any point on the surface. 
It is known for the real scalar models in three and four dimensions 
that the effective potential is convex and has flat bottom 
in the broken phase \cite{convex}. 
We may choose any point in the flat region as a stable vacuum. 
It is likely that this fact supports the existence 
of a massless particle in the broken phase. 
We will continue our analysis with the fact in mind, that is, 
we find out the broken phase nonperturbatively 
by searching massless eigenvalues. 

\section{\bf BASIS FUNCTION REGULARIZATION}
\label{basis}
The first principle of RG
is to find a flow which gives the same physics. 
It is not easy to calculate 
the effective interaction part 
of the Bloch-Horowitz Hamiltonian directly 
for our practical purpose of nonperturbative renormalization 
and we have to avoid energy dependences of renormalization, 
then we consider another approach. 
In our framework, we can draw RG flows by calculating spectra. 
We regularize the LF Hamiltonian
with the basis function regularization 
and try to calculate the critical line and the critical surface. 

In general, we can express an arbitrary state in the Fock space such as
\begin{equation}
  |\Psi({\bf P}) \rangle =
  \sum_{n=1}^{N_{\rm TD}}
  \frac{1}{\sqrt{n!}}
  \int\left[\prod_{i=0}^n d^2k_i\right]
  \delta^2({\bf P}-\sum_{i=1}^n {\bf k}_i)
  \psi_n({\bf k}_1,{\bf k}_2,\dots,{\bf k}_n)
  \prod_{i=0}^n a^\dagger({\bf k}_i)
  |0\rangle, 
\end{equation}
where ${\bf P}$ is the total momentum of the state and
the wavefunction $\psi_n$ is symmetric 
under exchanges of arbitrary two momenta. 
The limit $N_{\rm TD}\to\infty$ corresponds to the full theory
and the state is normalized as 
\begin{equation}
  \label{nor1}
  \langle \Psi({\bf P}) | \Psi({\bf Q}) \rangle
  = \delta^2({\bf P} - {\bf Q}).
\end{equation}
We will set a certain small number to $N_{\rm TD}$, 
which is called the Tamm-Dancoff approximation. 
In this paper, 
the Fock space is truncated up to three body states 
($N_{\rm TD}=3$). 
According to the variational principle, 
the mass spectrum in the truncated Fock space is given 
by solving (diagonalizing) the ES equation 
\begin{equation}
  \label{es}
  P^2|\Psi({\bf P}) \rangle =M^2|\Psi({\bf P}) \rangle. 
\end{equation}

We introduce a cutoff $\Lambda$ and
show the present regularization scheme later. 
All quantities belonging to the model are measured in units of
the $\Lambda$ and attached tilde. 
In units of the cutoff, eigenmass and running parameters are 
\begin{equation}
  \tilde{M}       \equiv M/\Lambda, \quad
  \tilde{v}       \equiv v/\Lambda^{1/2}, \quad
  \tilde{r}       \equiv r/\Lambda^2, \quad
  \tilde{\lambda} \equiv \lambda/\Lambda, \quad
  \tilde{w}       \equiv w, 
\end{equation}
then 
\begin{equation}
  \label{rcpl}
  \tilde{g}_n = g_n/\Lambda^{(6-n)/2}.
\end{equation}
The $\phi^6$ interaction is marginal 
as shown in Eq. (\ref{rcpl}). 
We will check the marginal coupling dependence 
of the critical line in the next section. 
The momentum is also redefined as
\begin{equation}
  {\bf x}_i = (x_i,X_i) \equiv (k_i^+/P^+, k_{i\perp}/\Lambda), 
\end{equation}
where $0<x_i<1$ since $k_i^+ > 0$. 
In this rescaled notation, 
the normalization condition Eq. (\ref{nor1}) 
of the wavefunction becomes
\begin{equation}
  \sum_{n=1}^{N_{\rm TD}}
  \int \left[ \prod_{i=1}^n d{\bf x}_i \right]
  \delta^2(\tilde{{\bf P}} - \sum_{i=1}^n {\bf x}_i)
  |\tilde{\psi}_n({\bf x}_1,{\bf x}_2,\dots,{\bf x}_n)|^2 = 1, 
\end{equation}
where
\begin{equation}
  \tilde{\bf P} \equiv (1,\tilde{P}_\perp)
  = (1,P_\perp / \Lambda), 
\end{equation}
and the rescaled wavefunctions are 
\begin{equation}
  \tilde{\psi}_n({\bf x}_1,{\bf x}_2,\dots,{\bf x}_n)
  \equiv 
  \left( P^+ \Lambda \right)^{\frac{n-1}{2}}
  \psi_n({\bf k}_1,{\bf k}_2,\dots,{\bf k}_n). 
\end{equation}
The ES equation in the rescaled notation is given in
appendix \ref{app_es}. 
It can be seen that the ES equation does not depend on 
the longitudinal momentum $P^+$ and the cutoff $\Lambda$. 

Regularization of transverse momenta and discretization of 
Fock space are closely related to each other. 
The regularization of the Hamiltonian is realized 
as a boundary condition of the wavefunctions. 
That is, we solve this eigenvalue problem, 
keeping the constraint 
that the transverse component of $\tilde{\psi}_3$ is zero 
at edges ($X_i=-1,1$). 
The ES equation is regularized with the naive transverse cutoff, 
\begin{equation}
  -\Lambda < k_{i\perp} < \Lambda \leftrightarrow -1<X_i<1, 
\end{equation}
since the ultraviolet divergence 
is caused by the transverse loop integral. 
The Fock space is discretized by expanding the wavefunction
in terms of basis functions in the momentum space
($-1<X_i<1, \quad 0<x_i<1$), 
\begin{eqnarray}
  \label{bf}
  \tilde{\psi}_n({\bf x}_1,{\bf x}_2,\dots,{\bf x}_n)
  =
  \sum_{{\bf k}=0}^{N_{\rm L}(n)}
  \sum_{{\bf l}=0}^{N_{\rm T}(n)} a^{(n)}_{\bf kl}
  {\cal S}_{1,2,\dots,n}
  \left[
  \prod_{i=1}^n
  f_{k_i}(x_i)
  F_{l_i}(X_i)
  \right], 
\end{eqnarray}
where
\begin{equation}
  \sum_{i=1}^n x_i = 1, \quad
  \sum_{i=1}^n X_i = \tilde{P}_\perp, 
\end{equation}
and ${\cal S}_{1,2,\dots,n}$ is a symmetrizer. 
The $N_{\rm L}(n)$ and $N_{\rm T}(n)$ 
are taken sufficiently large 
so that the eigenvalue of the ground state can converge. 
Longitudinal basis functions are
\begin{equation}
  f_k(x) =  x^{\beta(n) + k}, \quad
  0<x<1,
\end{equation}
for $0<\beta(n)<1$ and transverse basis functions are
\begin{equation}
  F_l(X) = (1-X^2)X^l, \quad
  -1<X<1. 
\end{equation}
The variational parameter $\beta(n)$ is tuned 
so that the ground state takes minimum eigenvalue. 
The behavior of the three-body wavefunction 
near the edges ($x=0,1$) is important
because of the kinetic term proper to LF\cite{berg,beta}. 
The coefficients $a^{(n)}_{\bf kl}$ are determined 
by diagonalizing the Hamiltonian 
in the momentum space spanned by the basis functions. 
Mathematically, 
Eq. (\ref{es}) has exact eigenvalues only in the case 
that the functional space is expanded in terms of
the complete set of basis functions. 
We can get only upper bounds of eigenvalues, 
since the wavefunction is expanded 
in terms of incomplete basis functions
in this calculation. 
However, we expect that the spectrum is described 
accurately for low-lying states, 
since shapes of the wavefunctions are simple;
for example, 
the calculated wavefunction of the ground state has no node. 

In this paper, 
a phase diagram which includes rescaling effects is evaluated, 
in accordance with Wilson's RG prescription\cite{wilson_kogut}. 
It is not needed to vary the cutoff 
because the transformed cutoff 
is rescaled to the original one in this RG. 
Parameter sets which give the same eigenvalue are calculated 
for a fixed cutoff. 

The wavefunction renormalization is neglected in this calculation, 
i.e., $Z_{\varphi}=1$. 
The phase diagram is dominated by the Gaussian fixed point, 
since the field is rescaled according to the canonical dimension. 
In order to find a non-trivial fixed point, 
we have to consider the wavefunction renormalization. 
These two phase diagrams, which have trivial and non-trivial fixed points, 
describe different theories. 
Our result makes sense as the theory dominated by the Gaussian fixed point. 

In this calculation, 
the total transverse momentum of the eigenstate is set to zero, 
$\tilde{P}_\perp = 0$, 
and $N_{\rm L,T}(n)=3$ in Eq. (\ref{bf}), 
which is sufficiently large so as to give a convergent spectrum. 

\section{\bf NUMERICAL RESULTS}
\label{numerical}
\begin{figure}[h]
\begin{center} \leavevmode
\epsfile{file=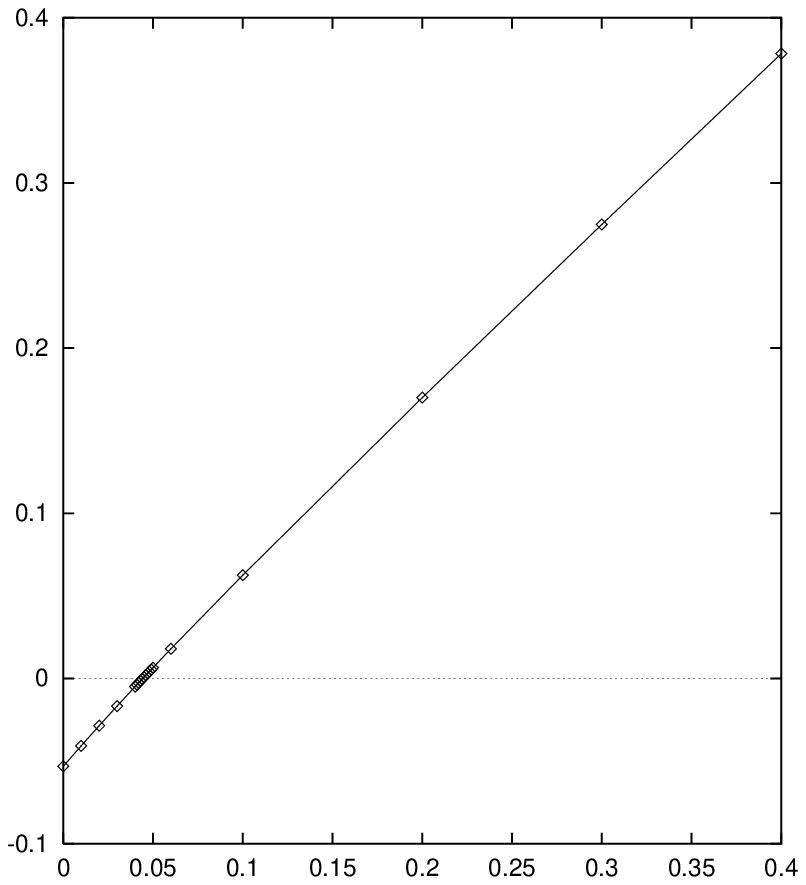,scale=1.2}
\put(-130,160){one-boson}
\put(-143,-20){$\tilde{r}$}
\put(-310,160){$\tilde{M}^2$}
\put(-255,280){$\tilde{\lambda}=10$}
\put(-235,57){$\tilde{r}_{\rm c}(\tilde{\lambda}=10)$}
\end{center}
\caption{
Mass spectrum for the lowest eigenstate with $\tilde{\lambda}=10$. 
The spectrum is plotted 
as a function of the input mass parameter $\tilde{r}$. 
$\tilde{r}_{\rm c}$ is a critical point, 
which gives the massless ground state. 
}
\label{cpl10}
\end{figure}

\begin{figure}[h]
\begin{center} \leavevmode
\epsfile{file=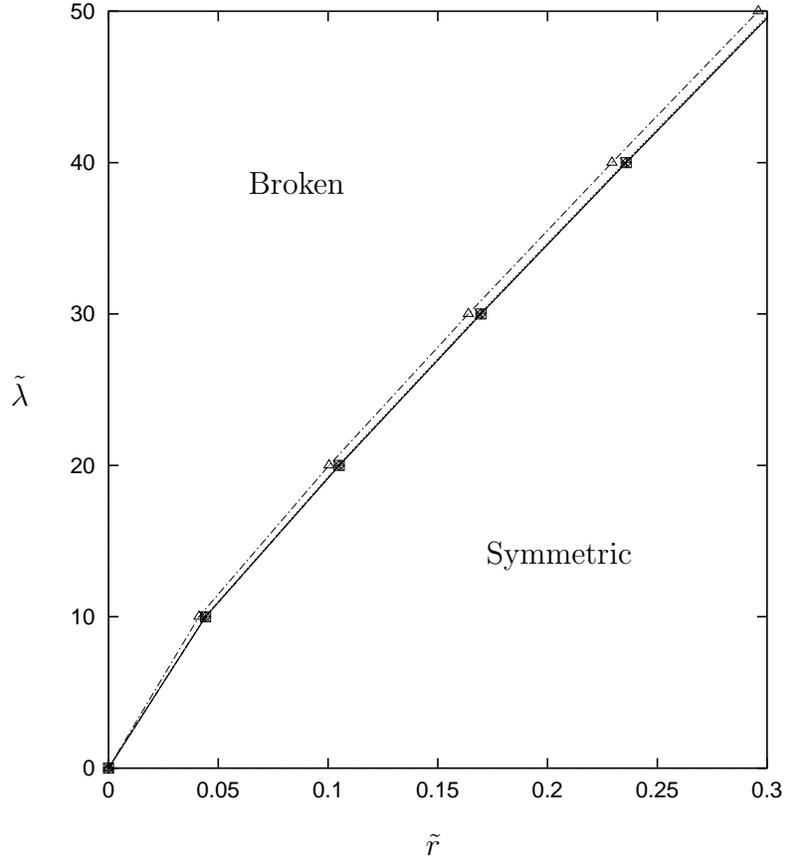,scale=1.2}
\put(-210,230){Broken}
\put(-120,90){Symmetric}
\put(-143,-20){$\tilde{r}$}
\put(-300,150){$\tilde{\lambda}$}
\end{center}
\caption{
Critical lines for various values 
of the marginal coupling constant $\tilde{w}$. 
The diamonds, bars, squares, crosses and triangles correspond to
$\tilde{w}=0$, $\tilde{w}=1$, $\tilde{w}=10$, 
$\tilde{w}=10^2$ and $\tilde{w}=10^3$, respectively. 
The critical line little depends on the marginal coupling constant. 
}
\label{mcl}
\end{figure}

Mass spectrum is plotted in Fig. \ref{cpl10} 
for the case $\tilde{\lambda}=10$ and $\tilde{w}=0$
without introducing $\tilde{v}$. 
No multi-bosonic bound state is found in this calculation. 
We can observe a massless point $\tilde{r}_{\rm c}(\tilde{\lambda}=10)$. 
The point is called the critical point. 
We can draw the critical line by connecting the critical points 
which are calculated 
for various $\tilde{\lambda}$. 
The critical line is plotted in Fig. \ref{mcl} 
for five values of marginal coupling constants, 
$\tilde{w}=0, 1, 10, 10^2, 10^3$. 
Values of two parameters $\tilde{r}$ and $\tilde{\lambda}$ 
which form the critical lines
are tabulated in table \ref{critical} for each $\tilde{w}$.
The dependence of the line on $\tilde{w}$ is weak. 
This is consistent with our intuition 
based on perturbative analysis, i.e., 
the dependence of the line 
on higher power operators in $\varphi$ may be week. 
After this, we will switch off the marginal coupling ($\tilde{w}=0$), 
and draw the phase diagram of three running parameters, 
$\tilde{v}$, $\tilde{r}$ and $\tilde{\lambda}$
near the $\tilde{v}=0$ plane. 

\begin{figure}[h]
\begin{center} \leavevmode
\epsfile{file=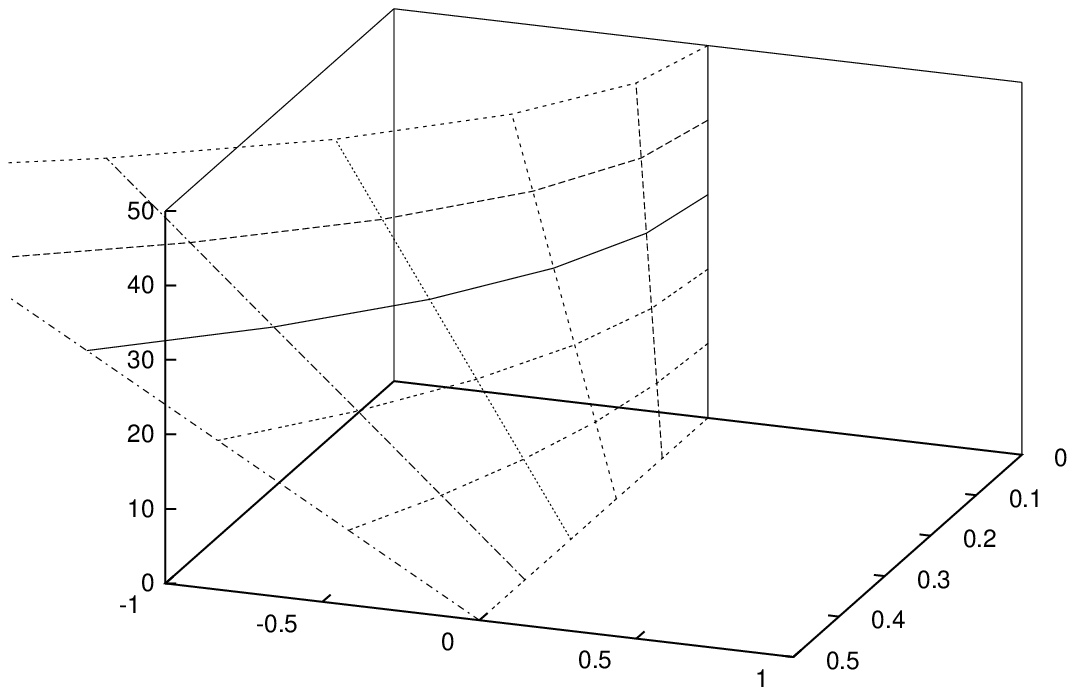,scale=0.9}
\put(-296,100){$\tilde{\lambda}$}
\put(-200,16){$\tilde{r}$}
\put(-65,32){$\tilde{v}$}
\put(-270,185){(a)}
\end{center}
%
\begin{center}\leavevmode
\epsfile{file=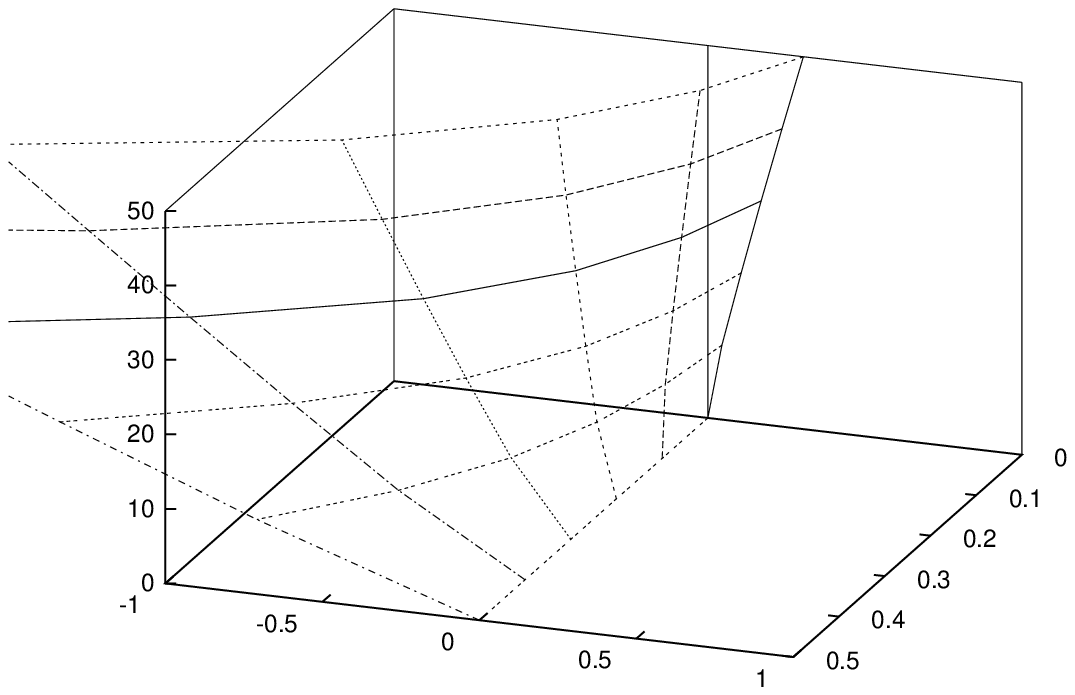,scale=0.9}
\put(-296,100){$\tilde{\lambda}$}
\put(-200,16){$\tilde{r}$}
\put(-65,32){$\tilde{v}$}
\put(-270,185){(b)}
\end{center}
\caption{
(a)A surface for the relation 
$\tilde{\lambda}\tilde{v}^2+6\tilde{r}=0$
is figured. 
(b)A critical surface, which gives the massless ground state, 
is calculated by diagonalizing the light-front Hamiltonian 
in the Fock space truncated up to three-body states. 
}
\label{surface}
\end{figure}

The lowest eigenmass is zero on the line, 
corresponding to an infinite correlation length 
in the statistical theoretical language. 
The spectrum of the canonical Hamiltonian 
is not bounded from below 
in the left region of the critical line. 
We then conclude that 
excited states are lighter than the ground state in the region 
if we solve the canonical Hamiltonian as it is. 
This is a result of the mass renormalization effect 
explained in the section \ref{shift}. 
The true Hamiltonian should be positive definite 
even in the left region of the critical line. 
It is natural to introduce the VEV 
in the broken phase 
in order to construct a positive definite Hamiltonian. 
The symmetry has to be spontaneously broken
for the spectrum to keep positive. 
Figure \ref{surface}(a) shows a relation 
$\tilde{\lambda} \tilde{v}^2 + 6 \tilde{r} = 0$
among three parameters 
$\tilde{v}$, $\tilde{r}$ and $\tilde{\lambda}$, 
which is given by perturbative RG calculations at one-loop order; 
the detail of the calculations is shown 
in \cite{sugi} 
for the case of the four-dimensional scalar model.   
We can see a critical line on the positive $\tilde{\lambda}$ axis, 
which is trivial because no renormalization effect is included. 
In Fig. \ref{surface}(b), a critical surface is drawn. 
The surface is calculated by searching points 
which give massless spectrum 
in the three-dimensional parameter space 
$(\tilde{v},\tilde{r},\tilde{\lambda})$. 
Positivity of the Hamiltonian is restored 
by introducing $\tilde{v}$. 
Compared to the surface (a) in Fig. \ref{surface}, 
the surface (b) slants to the positive $\tilde{r}$ direction
if $\tilde{v}$ is small 
and to the negative direction if $\tilde{v}$ is large, 
as a result of renormalization effects. 

\begin{figure}[h]
\begin{center}\leavevmode
\epsfile{file=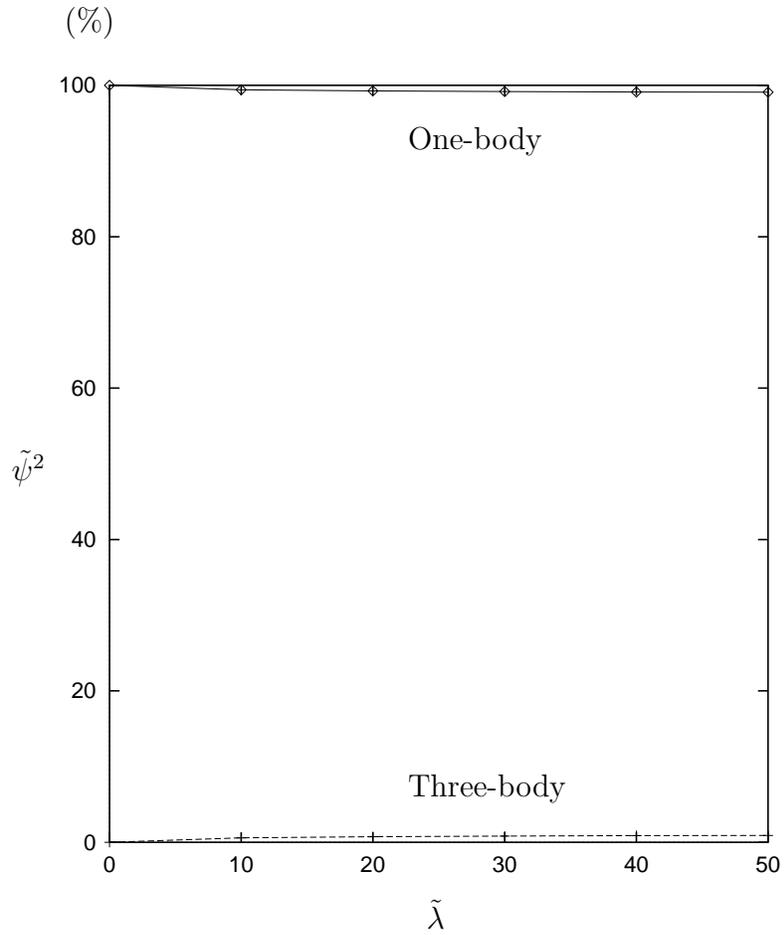,scale=1.2}
\put(-280,320){$(\%)$}
\put(-143,-20){$\tilde{\lambda}$}
\put(-300,150){$\tilde{\psi}^2$}
\put(-150,275){One-body}
\put(-150,30){Three-body}
\end{center}
\caption{
Wavefunction components of the ground state on the critical line;
one- and three-body components are shown by diamonds and crosses
as a function of $\tilde{\lambda}$, respectively. 
Lines are intended to guide the eyes. 
}
\label{ratio}
\end{figure}

Components of the ground state wavefunction on the critical line 
shown in Fig. \ref{mcl} are plotted in Fig. \ref{ratio} 
in order to confirm whether the TD approximation works well or not. 
One- and three-body wavefunction probabilities are
plotted as a function of $\tilde{\lambda}$ on the critical line. 
The three-body component is very small
even for comparatively strong coupling $\tilde{\lambda}$. 
This means that
the spectrum does not change near the critical line 
even if higher Fock space contributions are included in the calculation. 
This is true also in region far from the critical line, 
when the VEV is zero. 
The three-body component of the ground state wavefunction
tends to decrease, 
as mass parameter increases with coupling constant fixed, 
because taking large mass parameter 
is effectively the same as taking weak coupling constant. 

\begin{figure}[h]
\begin{center}\leavevmode
\epsfile{file=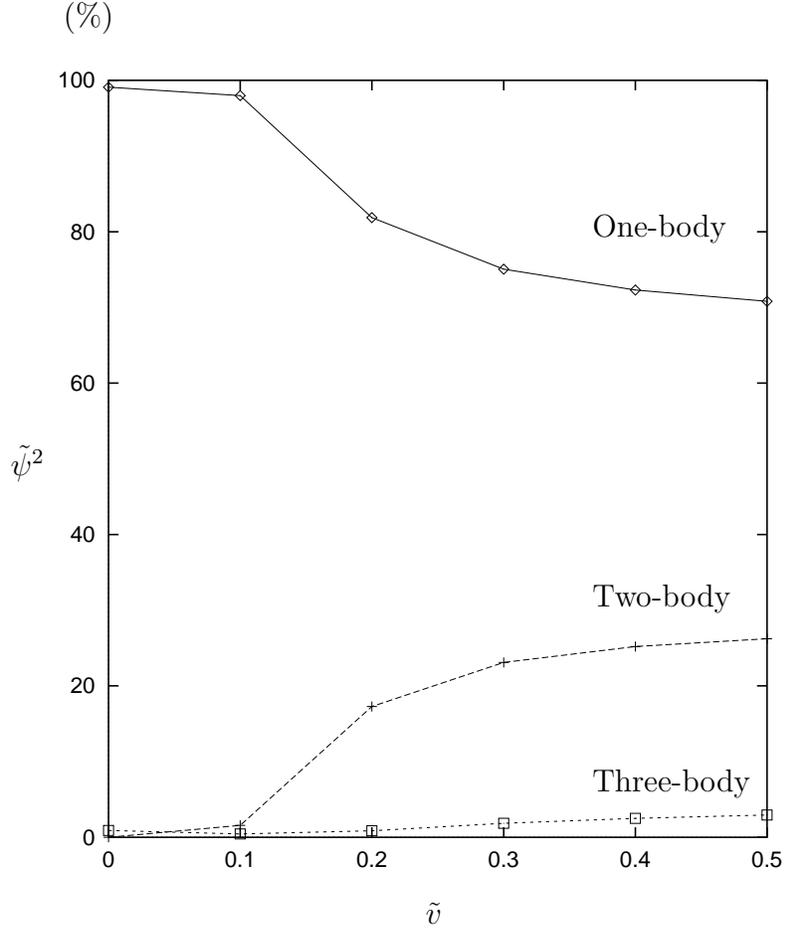,scale=1.2}
\put(-280,320){$(\%)$}
\put(-143,-20){$\tilde{v}$}
\put(-300,150){$\tilde{\psi}^2$}
\put(-80,240){One-body}
\put(-80,100){Two-body}
\put(-80,30){Three-body}
\end{center}
\caption{
Wavefunction components of the ground state on the critical line;
one-, two- and three-body components 
on an intersection between 
the critical surface and the $\tilde{\lambda}=50$ plane
are shown by diamonds, crosses and squares
as a function of $\tilde{v}$, respectively. 
Lines are intended to guide the eyes. 
}
\label{vac}
\end{figure}

Next, we investigate TD dependence of the critical surface. 
Figure \ref{vac} shows wavefunction components of the ground state 
on the intersection between the critical surface 
and the $\tilde{\lambda}=50$ plane. 
The coupling constant is set to the largest value in this calculation. 
One- and two-body components are dominant and
three-body component is small also in this case. 
The two-body component increases rapidly near $\tilde{v}=0.1$, 
whereas the three-body component increases slowly 
and keeps values around a few percent. 
It is expected near the $\tilde{v}=0$ plane that 
the Tamm-Dancoff approximation is good. 

\section{\bf SUMMARY AND DISCUSSIONS}
\label{discussion}
We have shown how to renormalize the Hamiltonian nonperturbatively 
for the three-dimensional real scalar model 
and calculated the critical line and the critical surface. 
There exist a region 
where the canonical Hamiltonian $H_{\rm S}$ is unstable 
independently of perturbation. 
The mass spectrum is tachyonic 
in the left region of the critical line. 
We have then introduced VEV 
to restore instability of the Hamiltonian
and calculated the critical surface. 
The asymmetric Hamiltonian $H_{\rm A}$ 
on the surface describes physics for the broken phase, 
if the phase has a massless mode. 
This is precisely true for SSB of any continuous symmetry, and 
the present approach for finding the broken phase is applicable 
for the case. 
Similarly to the present bosonic case, 
there seems to exist a region where Hamiltonian is unstable
if we renormalize a mass term $m\bar{\psi}\psi$
in fermionic theories. 
It is reasonable to understand that
the instability is restored by introducing VEV of a bosonic field. 

The marginal coupling dependence of the critical line is weak. 
This result links to the fact that 
the Tamm-Dancoff approximation up to three-body state works well. 
The ES equation for $\tilde{\psi}_1$ has no marginal interaction, 
but the ES equation for $\tilde{\psi}_3$ has that. 
The ground state has marginal coupling dependence 
only through the three-body wavefunction $\tilde{\psi}_3$ 
which is extremely small as a result of nonperturbative calculation. 
This is the reason 
why the marginal coupling dependence of the ground state is weak. 

This RG program seems to be independent of energy eigenvalues. 
In order to check this, we need plural number of bound states. 
For example, 
we can use spin singlet and degenerate triplet states 
in a fermionic theory such as Yukawa model. 
It is possible to draw RG flows 
by using a spectrum of one-body fermionic state 
in a similar way which has been done in this paper. 
We can say that the RG is energy independent
if the spectra of the spin singlet and triplet states 
are renormalized with the same RG flow. 

\section{ACKNOWLEDGMENTS}
We would like to thank K. Harada and 
S. Tominaga for helpful discussions. 
\appendix
\section{OSCILLATOR EXPANSION OF THE FIELD}
\label{expansion}
The field $\varphi$ is expanded at $x^+ = 0$ with oscillators as
\begin{equation}
  \varphi(x)_{x^+=0}
  =
  \int_0^\infty dk^+
  \int_{-\infty}^\infty
  \frac{dk_\perp}{\sqrt{2k^+ (2\pi)^2}}
  \left[
    a({\bf k})
    e^{-ik^+ x^- + ik_\perp x_\perp}
    +
    a^\dagger({\bf k})
    e^{ik^+ x^- - ik_\perp x_\perp}
  \right], 
\end{equation}
where
\begin{equation}
  \left[
    a({\bf k}), 
    a^\dagger({\bf k}')
  \right]
  =
  \delta^2({\bf k}-{\bf k}'). 
\end{equation}

\section{EINSTEIN SCHR\"ODINGER EQUATION}
\label{app_es}
The ES equation (\ref{es}) becomes
\begin{eqnarray}
  \label{sch}
  &&\tilde{M}^2
  \tilde{\psi}_n({\bf x}_1,{\bf x}_2,\dots,{\bf x}_n) \\
  \nonumber
  &=&
  \left[
    \sum_{i=1}^n \tilde{\epsilon}({\bf x}_i) - \tilde{\bf P}_\perp^2
  \right]
  \tilde{\psi}_n({\bf x}_1,{\bf x}_2,\dots,{\bf x}_n) \\
  \nonumber
  &+&
  \frac{\tilde{g}_4}{4!}
  \frac{6}{(2\pi)^2}
  \int_{\bf y}
  \sum_{i<j}^n
  \frac{\delta^2({\bf x}_i+{\bf x}_j - \sum_{i=1}^2 {\bf y}_i)}
  {\sqrt{x_i x_j}}
  \\
  \nonumber
  && \hspace{4.5cm}\times
  \tilde{\psi}_n(
  {\bf y}_1,
  {\bf y}_2,
  {\bf x}_1,
  \dots,
  \check{\bf x}_i,
  \dots,
  \check{\bf x}_j,
  \dots,
  {\bf x}_n) \\
  \nonumber
  &+&
  \frac{\tilde{g}_6}{6!}
  \frac{30}{(2\pi)^4}
  \int_{\bf y}
  \sum_{i<j<k}^n
  \frac{\delta^2({\bf x}_i+{\bf x}_j+{\bf x}_k - \sum_{i=1}^3 {\bf y}_i)}
  {\sqrt{x_i x_j x_k}}
  \\
  \nonumber
  && \hspace{4.5cm}\times
  \tilde{\psi}_n(
  {\bf y}_1,
  {\bf y}_2,
  {\bf y}_3,
  {\bf x}_1,
  \dots,
  \check{\bf x}_i,
  \dots,
  \check{\bf x}_j,
  \dots,
  \check{\bf x}_k,
  \dots,
  {\bf x}_n) \\
  \nonumber
  &+&
  \frac{\tilde{g}_3}{3!}
  \frac{3}{\sqrt{2}(2\pi)}
  \sqrt{\frac{(n+1)!}{n!}}
  \int_{\bf y}
  \sum_{i=1}^n
  \frac{\delta^2({\bf x}_i - \sum_{i=1}^2 {\bf y}_i)}
  {\sqrt{x_i}}
  \\
  \nonumber
  && \hspace{4.5cm}\times
  \tilde{\psi}_{n+1}(
  {\bf y}_1,
  {\bf y}_2,
  {\bf x}_1,
  \dots,
  \check{\bf x}_i,
  \dots,
  {\bf x}_n) \\
  \nonumber
  &+&
  \frac{\tilde{g}_5}{5!}
  \frac{10}{\sqrt{2}(2\pi)^3}
  \sqrt{\frac{(n+1)!}{n!}}
  \int_{\bf y}
  \sum_{i<j}^n
  \frac{\delta^2({\bf x}_i+{\bf x}_j - \sum_{i=1}^3 {\bf y}_i)}
  {\sqrt{x_i x_j}}
  \\
  \nonumber
  && \hspace{4.5cm}\times
  \tilde{\psi}_{n+1}(
  {\bf y}_1,
  {\bf y}_2,
  {\bf y}_3,
  {\bf x}_1,
  \dots,
  \check{\bf x}_i,
  \dots,
  \check{\bf x}_j,
  \dots,
  {\bf x}_n) \\
  \nonumber
  &+&
  \frac{\tilde{g}_3}{3!}
  \frac{6}{\sqrt{2}(2\pi)}
  \sqrt{\frac{(n-1)!}{n!}}
  \int_{\bf y}
  \sum_{i<j}^n
  \frac{\delta^2({\bf x}_i+{\bf x}_j - {\bf y})}
  {\sqrt{x_i x_j}}
  \\
  \nonumber
  && \hspace{4.5cm}\times
  \tilde{\psi}_{n-1}(
  {\bf y},
  {\bf x}_1,
  \dots,
  \check{\bf x}_i,
  \dots,
  \check{\bf x}_j,
  \dots,
  {\bf x}_n) \\
  \nonumber
  &+&
  \frac{\tilde{g}_5}{5!}
  \frac{30}{\sqrt{2}(2\pi)^3}
  \sqrt{\frac{(n-1)!}{n!}}
  \int_{\bf y}
  \sum_{i<j<k}^n
  \frac{\delta^2({\bf x}_i+{\bf x}_j+{\bf x}_k - {\bf y})}
  {\sqrt{x_i x_j x_j}}
  \\
  \nonumber
  && \hspace{4.5cm}\times
  \tilde{\psi}_{n-1}(
  {\bf y}_1,
  {\bf y}_2,
  {\bf x}_1,
  \dots,
  \check{\bf x}_i,
  \dots,
  \check{\bf x}_j,
  \dots,
  \check{\bf x}_k,
  \dots,
  {\bf x}_n) \\
  \nonumber
  &+&
  \frac{\tilde{g}_4}{4!}
  \frac{2}{(2\pi)^2}
  \sqrt{\frac{(n+2)!}{n!}}
  \int_{\bf y}
  \sum_{i=1}^n
  \frac{\delta^2({\bf x}_i - \sum_{i=1}^3 {\bf y}_i)}
  {\sqrt{x_i}}
  \\
  \nonumber
  && \hspace{4.5cm}\times
  \tilde{\psi}_{n+2}(
  {\bf y}_1,
  {\bf y}_2,
  {\bf y}_3,
  {\bf x}_1,
  \dots,
  \check{\bf x}_i,
  \dots,
  {\bf x}_n) \\
  \nonumber
  &+&
  \frac{\tilde{g}_6}{6!}
  \frac{15}{2(2\pi)^4}
  \sqrt{\frac{(n+2)!}{n!}}
  \int_{\bf y}
  \sum_{i<j}^n
  \frac{\delta^2({\bf x}_i+{\bf x}_j - \sum_{i=1}^4 {\bf y}_i)}
  {\sqrt{x_i x_j}}
  \\
  \nonumber
  && \hspace{4.5cm}\times
  \tilde{\psi}_{n+2}(
  {\bf y}_1,
  {\bf y}_2,
  {\bf y}_3,
  {\bf y}_4,
  {\bf x}_1,
  \dots,
  \check{\bf x}_i,
  \dots,
  \check{\bf x}_j,
  \dots,
  {\bf x}_n) \\
  \nonumber
  &+&
  \frac{\tilde{g}_4}{4!}
  \frac{12}{(2\pi)^2}
  \sqrt{\frac{(n-2)!}{n!}}
  \int_{\bf y}
  \sum_{i<j<k}^n
  \frac{\delta^2({\bf x}_i+{\bf x}_j+{\bf x}_k - {\bf y})}
  {\sqrt{x_i x_j x_k}}
  \\
  \nonumber
  && \hspace{4.5cm}\times
  \tilde{\psi}_{n-2}(
  {\bf y},
  {\bf x}_1,
  \dots,
  \check{\bf x}_i,
  \dots,
  \check{\bf x}_j,
  \dots,
  \check{\bf x}_k,
  \dots,
  {\bf x}_n) \\
  \nonumber
  &+&
  \frac{\tilde{g}_6}{6!}
  \frac{90}{(2\pi)^4}
  \sqrt{\frac{(n-2)!}{n!}}
  \int_{\bf y}
  \sum_{i<j<k<l}^n
  \frac{\delta^2({\bf x}_i+{\bf x}_j+{\bf x}_k+{\bf x}_l -
    \sum_{i=1}^2 {\bf y}_i)}
  {\sqrt{x_i x_j x_k x_l}}
  \\
  \nonumber
  && \hspace{4.5cm}\times
  \tilde{\psi}_{n-2}(
  {\bf y}_1,
  {\bf y}_2,
  {\bf x}_1,
  \dots,
  \check{\bf x}_i,
  \dots,
  \check{\bf x}_j,
  \dots,
  \check{\bf x}_k,
  \dots,
  \check{\bf x}_l,
  \dots,
  {\bf x}_n) \\
  \nonumber
  &+&
  \frac{\tilde{g}_5}{5!}
  \frac{5}{2\sqrt{2}(2\pi)^3}
  \sqrt{\frac{(n+3)!}{n!}}
  \int_{\bf y}
  \sum_{i=1}^n
  \frac{\delta^2({\bf x}_i - \sum_{i=1}^4 {\bf y}_i)}
  {\sqrt{x_i}}
  \\
  \nonumber
  && \hspace{4.5cm}\times
  \tilde{\psi}_{n+3}(
  {\bf y}_1,
  {\bf y}_2,
  {\bf y}_3,
  {\bf y}_4,
  {\bf x}_1,
  \dots,
  \check{\bf x}_i,
  \dots,
  {\bf x}_n) \\
  \nonumber
  &+&
  \frac{\tilde{g}_5}{5!}
  \frac{60}{\sqrt{2}(2\pi)^3}
  \sqrt{\frac{(n+3)!}{n!}}
  \int_{\bf y}
  \sum_{i<j<k<l}^n
  \frac{\delta^2({\bf x}_i+{\bf x}_j+{\bf x}_k+{\bf x}_l - {\bf y})}
  {\sqrt{x_i x_j x_k x_l}}
  \\
  \nonumber
  && \hspace{4.5cm}\times
  \tilde{\psi}_{n-3}(
  {\bf y},
  {\bf x}_1,
  \dots,
  \check{\bf x}_i,
  \dots,
  \check{\bf x}_j,
  \dots,
  \check{\bf x}_k,
  \dots,
  \check{\bf x}_l,
  \dots,
  {\bf x}_n) \\
  \nonumber
  &+&
  \frac{\tilde{g}_6}{6!}
  \frac{3}{2(2\pi)^4}
  \sqrt{\frac{(n+4)!}{n!}}
  \int_{\bf y}
  \sum_{i=1}^n
  \frac{\delta^2({\bf x}_i - \sum_{i=1}^5 {\bf y}_i)}
  {\sqrt{x_i}}
  \\
  \nonumber
  && \hspace{4.5cm}\times
  \tilde{\psi}_{n+4}(
  {\bf y}_1,
  {\bf y}_2,
  {\bf y}_3,
  {\bf y}_4,
  {\bf y}_5,
  {\bf x}_1,
  \dots,
  \check{\bf x}_i,
  \dots,
  {\bf x}_n) \\
  \nonumber
  &+&
  \frac{\tilde{g}_6}{6!}
  \frac{180}{(2\pi)^4}
  \sqrt{\frac{(n-4)!}{n!}}
  \int_{\bf y}
  \sum_{i<j<k<l<m}^n
  \frac{\delta^2({\bf x}_i+{\bf x}_j+{\bf x}_k+{\bf x}_l+{\bf x}_m - {\bf y})}
  {\sqrt{x_i x_j x_k x_l x_m}}
  \\
  \nonumber
  && \hspace{4.5cm}\times
  \tilde{\psi}_{n-4}(
  {\bf y},
  {\bf x}_1,
  \dots,
  \check{\bf x}_i,
  \dots,
  \check{\bf x}_j,
  \dots,
  \check{\bf x}_k,
  \dots,
  \check{\bf x}_l,
  \dots,
  \check{\bf x}_m,
  \dots,
  {\bf x}_n), 
  \nonumber
\end{eqnarray}
where $\tilde{\epsilon}({\bf x}) \equiv (X^2+g_2)/x$ and 
\begin{equation}
  \int_{\bf y}
  \equiv
  \int
  \left[
  \prod_{i=1}^n
  \frac{d{\bf y}_i}{\sqrt{y}_i} 
  \right], 
\end{equation}
this shows integrations over $n$-pieces of ${\bf y}$, 
and the ${\bf x}$'s with the check symbol  
are removed from arguments of the wavefunctions. 
On the right hand side of Eq. (\ref{sch}), 
interaction terms in which the number of checked $x$'s is 
larger than the number of particles $n$
are switched off. 
%
%

%
\vfill\eject
%
\vfill\eject
%
\vfill\eject
%
\vfill\eject
%
%
%
\begin{table}
\caption{
Values of mass parameter $\tilde{r}$ 
which give massless modes on the $\tilde{v}=0$ plane 
are tabulated as a function of $\tilde{\lambda}$ 
in the cases 
$\tilde{w}=0$, $\tilde{w}=1$, $\tilde{w}=10$, $\tilde{w}=10^2$ 
and $\tilde{w}=10^3$. 
The marginal coupling dependence of the critical line 
is weak. 
}
\begin{tabular}{d|dddddd}
  $\tilde{w}\backslash \tilde{\lambda}$
  & 0.0 & 10.0  & 20.0  & 30.0  & 40.0 & 50.0 \\
  \tableline
  0.0    & 0.000 & 0.044 & 0.105 & 0.170 & 0.236 & 0.303 \\
  $10^0$ & 0.000 & 0.044 & 0.105 & 0.170 & 0.236 & 0.303 \\
  $10^1$ & 0.000 & 0.044 & 0.105 & 0.170 & 0.236 & 0.303 \\
  $10^2$ & 0.000 & 0.044 & 0.105 & 0.170 & 0.235 & 0.302 \\
  $10^3$ & 0.000 & 0.041 & 0.100 & 0.164 & 0.230 & 0.296 \\
\end{tabular}
\label{critical}
\end{table}
\end{document}